# Light commodity devices for building vehicular ad hoc networks: An experimental study

Jamal Toutouh[a,*], Enrique Alba[a,b]

[a] *Departamento de Lenguajes y Ciencias de la Computación, Universidad de Málaga, Málaga, Spain*
[b] *VSB-Technical University of Ostrava, Ostrava, Czech Republic*

ABSTRACT

Vehicular communication networks represent both an opportunity and a challenge for providing smart mobility services by using a hybrid solution that relies on cellular connectivity and short range communications. The evaluation of this kind of network is overwhelmingly carried out in the present literature with simulations. However, the degree of realism of the results obtained is limited because simulations simplify real world interactions too much in many cases. In this article, we define an outdoor testbed to evaluate the performance of short range vehicular communications by using real world personal portable devices (smartphones, tablets, and laptops), two different PHY standards (IEEE 802.11g and IEEE 802.11a), and vehicles. Our test results on the 2.4 GHz band show that smartphones can be used to communicate vehicles within a range up to 75 m, while tablets can attain up to 125 m in mobility conditions. Moreover, we observe that vehicles equipped with laptops exchange multimedia information with nodes located further than 150 m. The communications on the 5 GHz band achieved an effective transmission range of up to 100 m. This, together with the optimization of the protocols used, could take our commodity lightweight devices to a new realm of use in the next generation of ad hoc mobility communications for moving through the city.

## 1. Introduction

Today, the longstanding promise of deploying applications to improve efficiency and safety in road transport is becoming a reality. A number of *smart mobility* solutions based on Intelligent Transport Systems (ITS) provide road users with information of traffic conditions by using wireless communication. This information is generally delivered to in-vehicle navigation systems via FM radio broadcast by using the Traffic Message Channel (TMC) or via digital media by using the Transport Protocol Experts Group (TPEG) [1]. New applications for smartphones and tablets may also receive this information via 3G or Long Term Evolution (LTE, 4G) cellular networks.

However, there are several drawbacks to most of these services [2]: (i) they are centralized and based on a fixed and costly infrastructure, e.g., over-roadway and in-roadway sensors; (ii) such systems only provide traffic information about the main roads in the city; and (iii) the information updates are in the range of 20–50 min, far from ideal real-time.

Over the last decade, the research community, the industry, and the authorities have been working on the deployment of vehicular ad hoc networks (VANETs). The idea is to equip vehicles with devices with wireless capabilities, such as dedicated on-board units or general-purpose smartphones and tablets, that allow short range communications to deploy volatile wireless networks on the roads. Thus, vehicles, as VANET nodes, may exchange useful traffic information to each other by using vehicle-to-vehicle (V2V) communications. In turn, infrastructure elements, i.e., traffic lights, traffic signs, information panels, etc., with wireless communication capabilities are other types of VANET nodes that communicate with vehicles by using vehicle-to-infrastructure (V2I) communications. Recently, the whole system has been extended to include personal pedestrian devices (P2I) as VANET nodes (see Fig. 1).

* Corresponding author. Tel.: +34 952 133303.
*E-mail addresses:* jamal@lcc.uma.es (J. Toutouh), eat@lcc.uma.es (E. Alba).

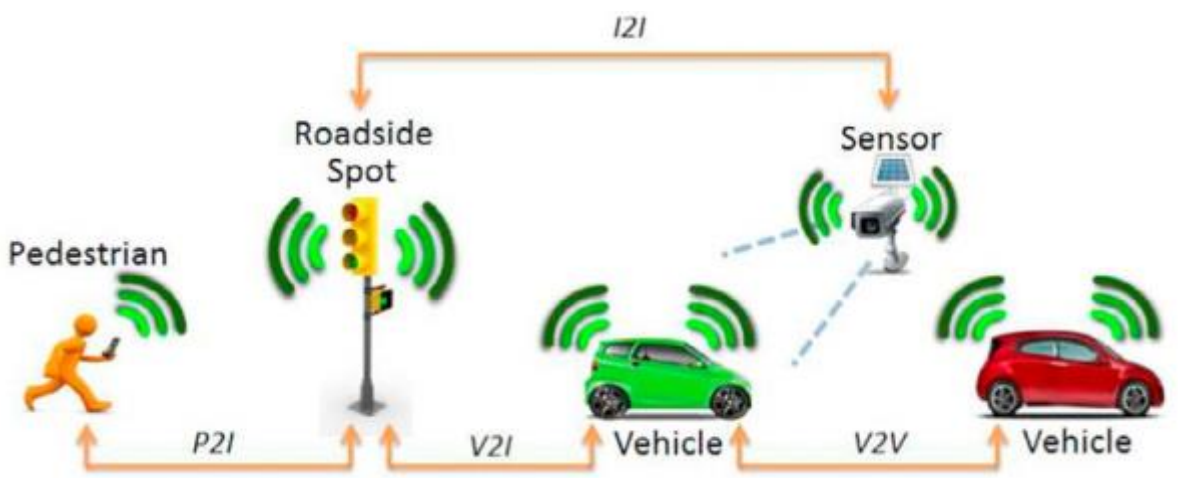

**Fig. 1.** Communication scheme between nodes in a VANET.

VANET technologies represent an opportunity to improve ITS services and applications by using short range communications. The main advantages of using VANETs are that (1) they are decentralized and totally distributed networks, (2) the vehicles are equipped with sensors that generate and update data about all roads visited, and (3) the information received can be fully personalized according to the user. However, it should be taken into account that VANET's performance is highly dependent on its market penetration. The more vehicles in the VANETs (distributed over more roads), the better the service they provide (more information and more updates).

Currently, much of the research effort is focused on defining the most appropriate architecture for vehicular networks (radio technologies, protocols, etc.) and on developing the promising ITS services to achieve smart mobility [3]. This research has resulted in three major architectures for vehicular communications: CALM (communications access for land mobiles) proposed by the International Standard Organization (ISO) [4], WAVE (wireless access in vehicular environments) produced by the Institute of Electrical and Electronic Engineers (IEEE) [5], and ITS-5G fostered by the European Telecommunications Standards Institute (ETSI) [6]. An important issue in this kind of research is the accurate evaluation of the solutions produced. As in most network environments, modelers, simulators, emulators, and application-dependent implementations are used for this purpose.

Modelers and simulators rely on mathematical formulas to determine the behavior of network protocols and applications. These analytic methods are limited by the complexity and dimension of real world systems, which usually require simplifications and approximations that generally lead to greater differences between their results and real world behavior. Emulators mimic the entire functionality of other systems (hardware, software, and network activity), accepting the same data, executing the same programs, and achieving similar results as the system being imitated. In the literature, the majority of the tools used to evaluate VANETs are a kind of modeler or simulator [7,8].

As a useful complement (or even realistic substitute) for simulations we can use experimental real testbeds. Testbeds have important advantages with respect to the aforementioned approaches because the test can be carried out in a real world environment offering close-to-real or real performance, as well as revealing behavioral issues [9]. However, there is a lack of scientific articles that use outdoor experiments in the field of vehicular networks. The main reasons for this may be the unavailability of resources (vehicles and road equipment), the difficulties in doing field studies, and the accuracy of the performance analysis.

In the present article, we carry out a set of outdoor experiments to analyze VANET communications. The idea is to prove the feasibility of VANET short range communications when using smartphones, tablets, and laptops. Moreover, we analyze other kinds of characteristics of these devices that are useful for VANET applications, e.g., the human machine interface (HMI) provided. In this article we do not intend to define a platform for vehicular communications differ- ent from the ones proposed by the ISO or the ETSI. Instead, we want to analyze the possibility of using such widespread lightweight commodity devices to provide ITS services to improve road transport in nowadays scenarios, where specific VANET devices are not available to most road users. The main goals of our work are therefore:

- Analyzing the main features that personal mobile devices provide for deploying VANETs without having to acquire new equipment for vehicles.
- Defining an outdoor testbed in an urban area to evaluate vehicle wireless communications.
- Studying the wireless capabilities of the devices analyzed in order to discuss their use in the deployment of VANETs.

The remainder of the paper is organized as follows. Section 2 reviews the literature in VANET communications analysis. Section 3 presents the three types of devices ana- lyzed in this work. Section 4 provides the details about the urban VANET testbed definition and experimental settings. Section 5 shows the results and discusses the performance observed in the experimentation. Finally, conclusions and fu- ture work are presented in Section 6.

## 2. Related work

In the last decade, a number of research projects managed by governmental, academic, and industrial entities, such as CARLINK [10], Cooperative Vehicle-Infrastructure Systems (CVIS) [11], SAFESPOT Integrated [12], DRIVE C2X [13], and PRESERVE [14], have been pushing to enhance the advanced driver assistance systems (ADAS) to provide cooperative road traffic solutions. In turn, the main international standardization institutes are involved in the specification of common frameworks and architectures for vehicular communications. The main initiatives proposed are the following: CALM by the ISO, WAVE by the IEEE, and C2C-CC (Car-to-Car Communication Consortium) [15] and ITS-5G by the ETSI.

The evaluation of the different approaches (radio technologies, protocols, and network models) to deploy vehicular networks is a major concern in this research field. In the literature, most studies aimed at this task have applied simulators and modelers. The authors have used three different types of strategies to simulate VANETs: utilizing a well-known network simulator, such as Ns-2/Ns-3 [7] or OPNET [16], that allows users to define the movement of the nodes by generating realistic VANET traces with transportation simulators, such as SUMO [17] or VISSIM [18]; coupling vehicular traffic and wireless network in a single simulator, e.g., GrooveNet [19]; and the most promising approach, synchronizing the existing and validated traffic and network simulators by using some specific bridge software, such as TraNS [20] or iTetris [8].

Even if the simulators have achieved a high degree of realism and they obtain quite accurate results, the real world simplifications that they apply in their mathematical definitions still limit their performance. Thus, in the literature, we can find some studies that define real world testbeds to analyze the feasibility of cooperative communications (V2V and V2I), and to test different VANET solutions, as protocols and applications.

The feasibility is generally proven by performing different measurements of the communications on the road in realistic conditions. An evaluation of V2I communications by attempting to exchange data between vehicles and a set of access points dotted around the city of Boston in USA was presented in [21]. Different studies [22–24] have analyzed the feasibility of VANET communications by using vehicle gateways defined by laptops equipped with a IEEE 802.11bg wireless interface with an external Bluetooth-based GPS.

Moreover, the literature presents other testbeds that focus on analyzing highly-demanding applications and protocols over different types of traffic conditions to study the main factors that affect their quality-of-service (QoS). Thus, [23,25] analyzed the performance of content sharing over VANETs, in [26], the authors analyzed FleetNet's complete platform services. All these studies used specific hardware (on-board units or laptops) installed in the vehicles.

Recently, as the last proposals of the ISO, the ETSI, and the IEEE are based on IEEE 802.11p [27], some studies were presented to analyze the capabilities of such a MAC and PHY layer for vehicular communications. The measurement of the performance of IEEE 802.11p PHY standard was carried out via V2I communications on an Austrian highway [28]. An analysis of Non-Line-Of-Sight (NLOS) reception quality of IEEE 802.11p at intersections in urban areas of Munich was presented in [29]. As the availability of IEEE 802.11p devices is very limited at present, other authors have decided to carry out experiments using IEEE 802.11a, which uses the band (5 GHz) closest to the one used by IEEE 802.11p. Sánchez et al. [24] studied VARON, a multi-hop wireless vehicular communication protocol, configuring the devices with IEEE 802.11a.

Finally, some other studies have analyzed ITS services that require the use of cellular networks (3G/LTE). For example, [30] presents an Android application that evaluates the behavior of the drivers and indicates how to reduce fleet degradation, insurance policy cost, and fuel consumption. Similarly, [9] introduces DRIVE, a reconfigurable VANET platform that allows experimental analysis of vehicular communications over real world tests that include direct short range and cellular network connectivity.

The interest of the present study is that the methodology includes an analysis of widely used portable devices (smartphones and tablets) when used in ad hoc communications, i.e., the use of such devices to exchange data with nearby nodes in a vehicular environment. In this analysis, we evaluate the wireless capabilities, but also, we study other features of these devices that can be useful in vehicular applications. Our analysis is neutral, in the sense that we do not focus on any one application as many other previous approaches have, thereby hopefully reaching a broader community of researchers. The use of such widely used devices to deploy VANETs could ease a faster penetration of this technology in our society.

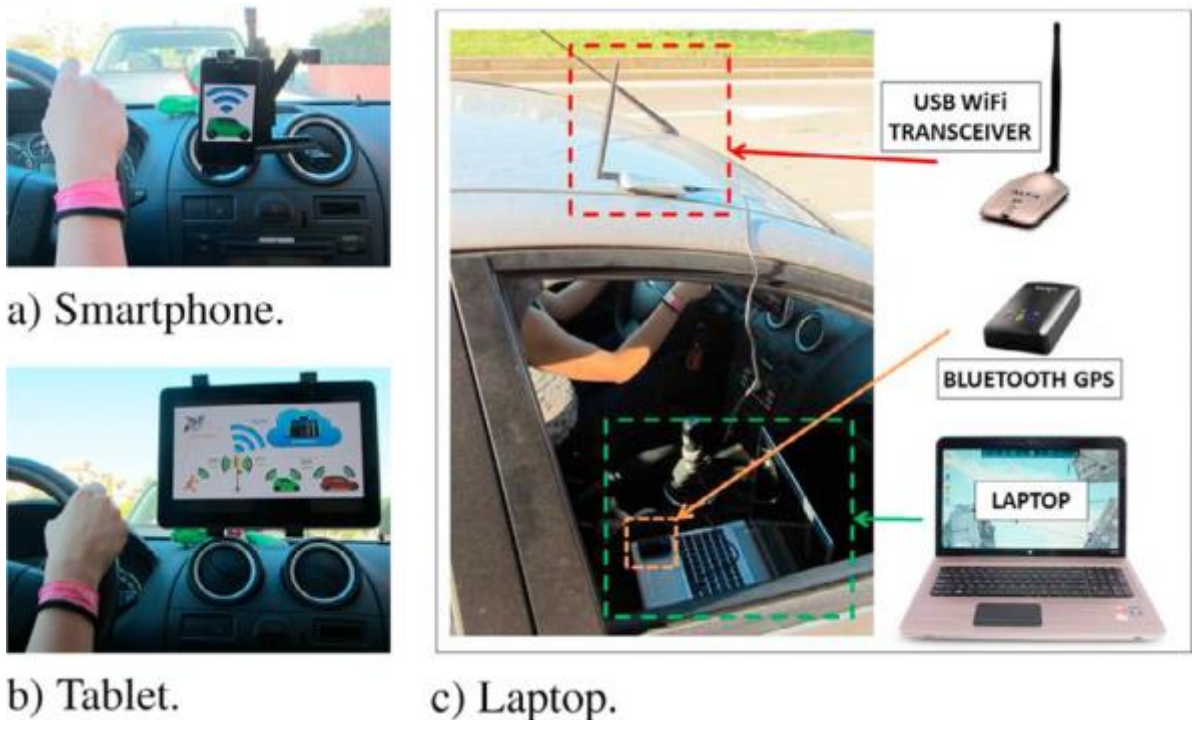

**Fig. 2.** Vehicle equipment used to perform the data transfers.

## 3. VANET ubiquitous devices

In our testbed, three different kinds of VANET nodes are defined depending on the type of the device used to equip the vehicles: a smartphone, a tablet or a laptop. To this end we use three devices that are readily available in the present market, as well as fairly standard in the features we are evaluating here: *Samsung Galaxy SII (Gt - I9100)* smartphones, *Samsung Galaxy Tab Gt - P7510* tablets, and *HP Pavilion dv6-3181ss* laptops that use *ALFA AWVS051NH* USB Wi-Fi transceivers. Fig. 2 shows our experimental VANET nodes.

The possible use of smartphones and tablets to deploy VANETs could accelerate the development of these kinds of networks, as these devices are already widely used by the population, this would in turn ease VANET's market penetration while the supply of on-board units specifically designed for vehicular communications is not widespread. In turn, these devices themselves are equipped with high processing capabilities (multi-core CPUs), many useful sensors for ITS applications (GPS, accelerometer, thermometer, compass, etc.), along with the required wireless connectivity (Bluetooth, Wi-Fi, and cellular links). Next, we present these three devices that have been used to deploy the VANET nodes of our testbed.

### 3.1. *Smartphones/Tablets for equipping VANET nodes*

Nowadays, our lives are flooded with personal mobile devices and it is common to see people walking around with a smartphone or a tablet in their hands. For this reason, we propose using them to deploy VANETs, thus avoiding the need to buy and install any new hardware in vehicles.

As shown in Fig. 3, smartphones and tablets provide a complete set of solutions to fulfill the requirements of a VANET node. In order to perform V2V/V2I, cellular, and in-vehicle communications they include IEEE 802.11bgn, cellular, and Bluetooth interfaces, respectively. The Bluetooth interface may be used to interact with the on-board diagnostics (OBD) reader and with the internal sensors of the vehicle. The internal GPS antenna makes the vehicle's geolocation possible, which is required by most ITS services and applications.

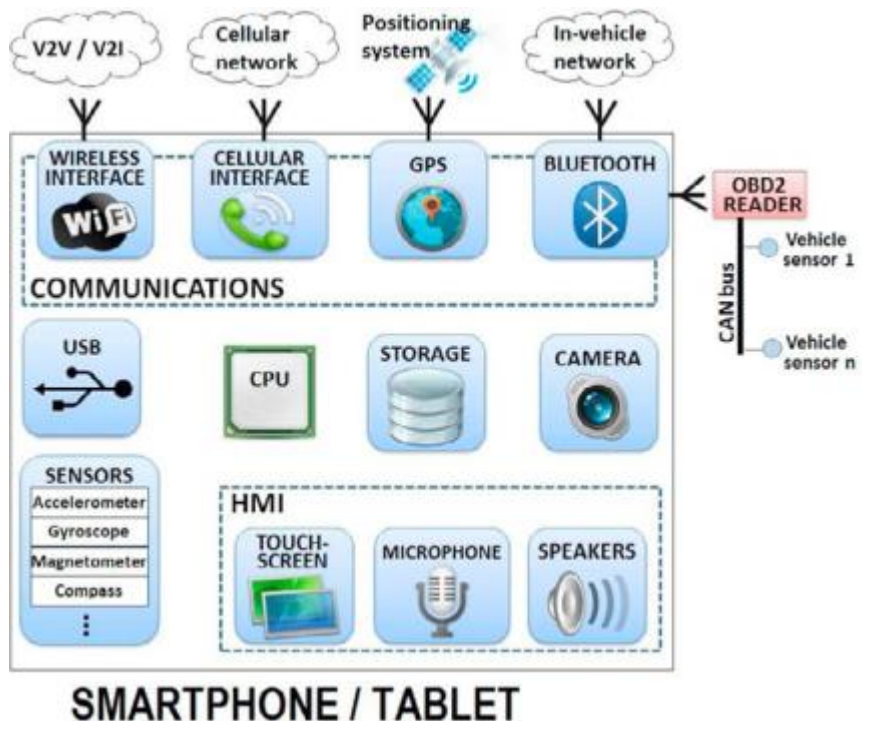

**Fig. 3.** Main features of the smartphones and tablets used in VANETs.

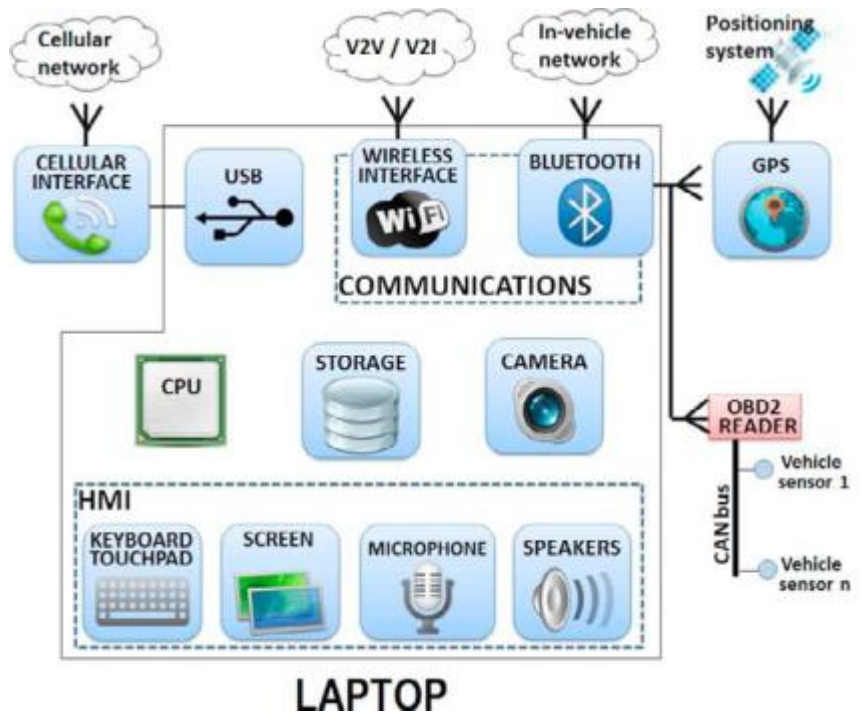

**Fig. 4.** Main features used in VANETs of laptops.

In addition, smartphones and tablets offer three different human-machine interfaces (HMI): the touch-screen, the microphone, and the speakers. The later two are really useful because drivers should not have to use their hands nor look away from the road to interact with the devices, thereby avoiding distractions that can cause dangerous situations on the road.

Both devices offer several internal hardware components such as one or two cameras, a USB interface, and a set of sensors. These components can be used by different types of services or applications, e.g., the accelerometer can be used to detect a collision to inform of possible road accidents.

In terms of wireless communication capabilities, both types of devices used here integrate the same *Broadcom BCM4330* wireless chip, but the tablet is bigger than the smartphone, and therefore, the antenna is larger and better located in tablet devices. Finally, in our testbed, we require the vehicles during the experiments to be located, and therefore, we utilize the GPS antenna integrated in these devices.

### 3.2. *Laptops for equipping VANET nodes*

By taking data when using laptops, we want to cover a range of scenarios where a powerful on-board unit is used. As shown in Fig. 4, most laptops provide two different types of wireless interfaces: IEEE 802.11abgn and Bluetooth. The first one can be used for V2V and V2I communications, and Bluetooth for in-vehicle communications. In order to perform cellular network communications the laptop requires an external cellular modem. In general, laptops do not include GPS antennas, so an external one may be required (see Fig. 2.c).

In the tests carried out in this study, we evaluate the V2V communications by using a laptop as the on-board unit. The laptops use an external *ALFA AWVS051NH* USB Wi-Fi transceiver, which includes an omnidirectional gain antenna that provides 2.5 dBi for transmitting on the 2.4 GHz band and 5 dBi for transmitting on the 5 GHz band. We also use the external Bluetooth GPS antenna to locate the vehicles during the experiments (see Fig. 2.c).

In terms of HMI, laptops usually provide a larger screen than smartphones and tablets, in addition to the keyboard, the touch-pad, the microphone, and the speakers. However, the idea is to use the speakers and the microphone to interact with the system to avoid possible distractions.

For the interaction with the vehicle, as in the case of mobile personal devices, a Bluetooth OBD reader can be used. It is important to take into account that smartphones and tablets provide a set of extra sensors, but laptops do not. So, ITS applications and services can access the information provided for the internal sensors installed in the car and the external sensors connected via USB or Bluetooth interfaces.

Finally, there is a drawback in using laptops as the main device for VANET nodes, which is the need for a specific space to place it in the car. This space is larger than the one needed to place smartphones or tablets.

## 4. Urban testbed definition

The experiments carried out in this study aim to analyze the feasibility of using personal mobile devices (smartphones and tablets) and laptops to deploy VANETs. Therefore, we have designed two different types of experiments: first, the evaluation of the power of the wireless signal generated by the analyzed devices, and second, tests to characterize the QoS of V2V communications when nodes use smartphones, tablets, and laptops.

### 4.1. *VANET scenarios*

The VANET testbed has been defined in an urban area of Málaga city (Spain) of 1.15 $Km^2$ covering a zone of the campus of the University of Málaga (see Fig. 5). The two experiments described here were both carried out in this area.

The evaluation of the power of the wireless signal generated by the devices studied has been carried out by measuring the signal strength at different points located at distances between 0 and 150 m. In this case the cars are stopped at a given point and we use another portable device to evaluate the power of the signal. As the cars are stopped we have named these experiments *static experiments*.

These *static experiments* have involved the signal strength evaluation of each device in 21 different points. The points are located at distances between 0 and 150 m and they are separated from each other by 7.5m (see Fig. 6.a). The distance

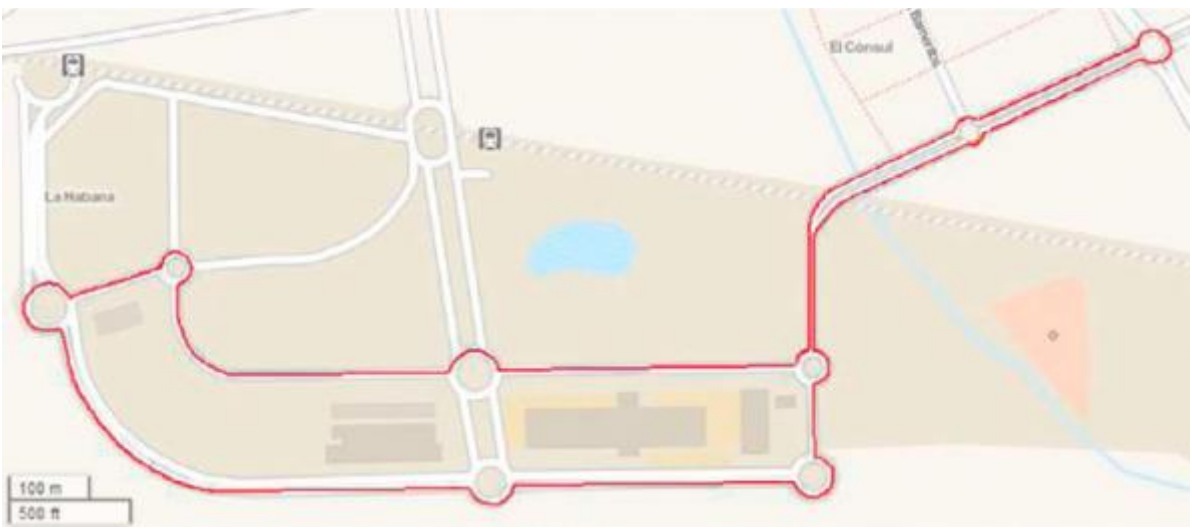

**Fig. 5.** Urban area where the testbed has been defined. The red line shows the circuit followed by the cars during the tests(For interpretation of the references to colour in this figure legend, the reader is referred to the web version of this article.).

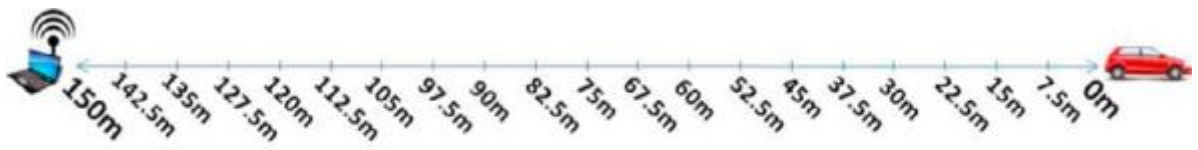

a) Distances at which the signal strength of the devices was measured when the nodes were not moving (static conditions).

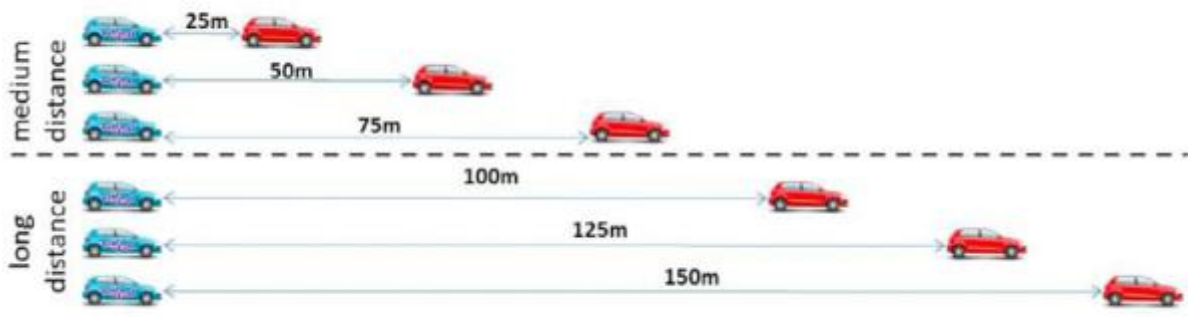

b) Distances between the cars during the V2V QoS analysis.

**Fig. 6.** Distances between the moving nodes during the experimentation.

of each point $i$ with the evaluated device is defined by (1).

$$distance_i = i \times 7.5\,m \;\; \forall i \in \{0, 1, \ldots, 20\} \quad (1)$$

The study of the feasibility of V2V communications has been done by defining a VANET comprising two cars that exchange data with each other. These cars move along a regular road of the testbed area that defines a circuit of 3,800 m (see Fig. 5). During the experiments, the speed of the vehicles is between 15 and 50 Km/h. Thus, we have evaluated different QoS metrics of the communications carried out in different situations. These experiments are called *dynamic experiments*.

In order to study how the distance between the VANET nodes also influences the performance of the communications, we have carried out different tests by modifying the distance between the vehicles over the journey in the *dynamic experiments*. As it was very difficult to maintain the same distances between the nodes as the ones defined for the experiments of the signal strength analysis because of the speed variations of the cars, we have used longer distances (distances multiple of 25 m). Thus, as shown in Fig. 6.b, we have defined different experiments with the two vehicles separated by six different distances grouped in *medium distance* (25, 50, and 75 m) and *long distance* (100, 125, and 150 m).

As the road was open to traffic, other vehicles were able to circulate, and therefore, our vehicles were affected as in a real life journey (neither constant speeds nor fixed distances between each other). In fact, other vehicles on the road could come in between the VANET nodes, thus affecting the wireless signal propagation (see Fig. 7.b). Note that in an urban area, buildings and other obstacles may also interfere and result in a far from ideal signal propagation: another reason supporting our choice of real world tests.

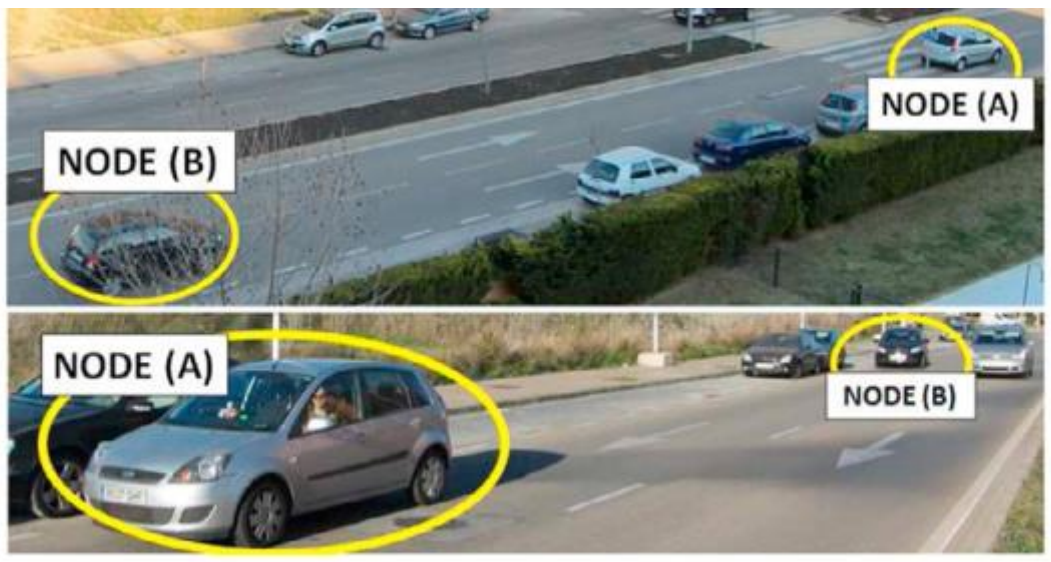

a) Cases with no vehicle inbetween our two moving cars.

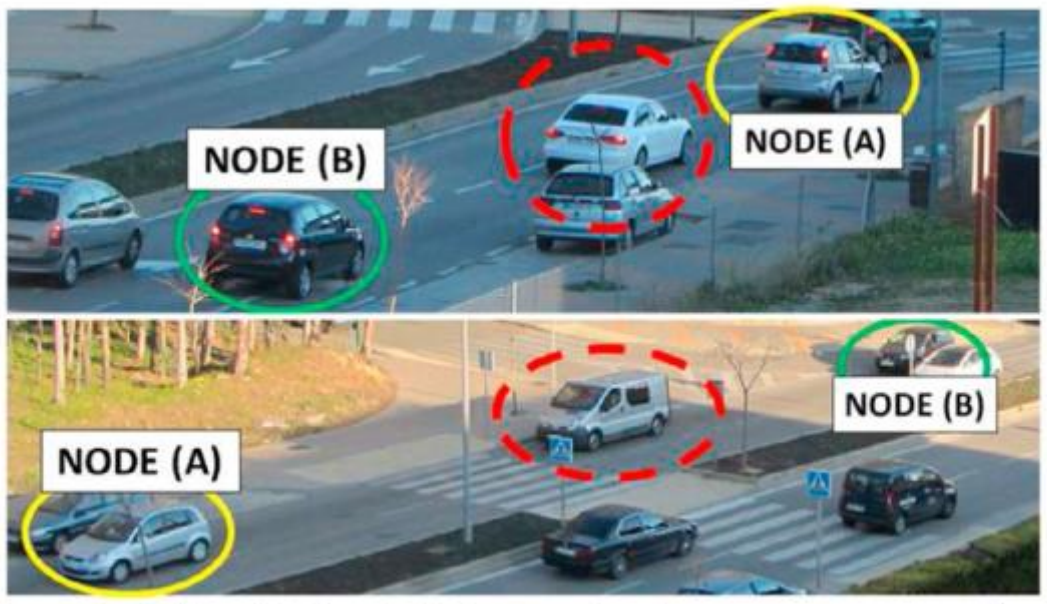

b) Cases of vehicles (in dotted circles) coming between our two moving cars.

**Fig. 7.** Nodes during the experiments. Having or not having cars in between the VANET nodes may change the communication behavior.

### 4.2. *Description of the tests*

The main idea is to analyze the use of lightweight devices' communication capabilities to deploy VANETs. Thus, we have selected the three different devices presented in Section 3 and defined two types of experiments: the wireless signal power measurement and the V2V data transfer evaluation.

The wireless interfaces of the smartphones, the tablets, and the laptops were configured with the widely used IEEE 802.11g mode that operates on the 2.4 GHz band and provides interoperability with IEEE 802.11b standard, and therefore, these devices can exchange information with most end user wireless devices. We also studied the communications between laptops when applying the IEEE 802.11a standard that operates on the 5 GHz band. We have also included this standard in our testbed because the vehicular communications proposed by the ISO, the ETSI, and the IEEE are all based on IEEE 802.11p, a modified version of this one [27].

Specifically, the devices use *Channel 6* of the 2.4 GHz band, whose nominal frequency is 2.437 GHz during IEEE 802.11g experiments. We have selected this channel because when channels 1, 6 or 11 are used, there is less potential for interference from other IEEE 802.11bg access points [31]. For IEEE 802.11a experiments, the devices use *Channel 140* of the 5 GHz band, whose nominal frequency is 5.700 GHz.

The *power of the signal* of each device is evaluated in *static experiments* which consist of evaluating the received signal strength in a given point for 15 s (see Section 4.1).

In order to evaluate the *feasibility* of V2V communications by using the three types of devices and the two IEEE 802.11 standards, we have defined *dynamic experiments*. These

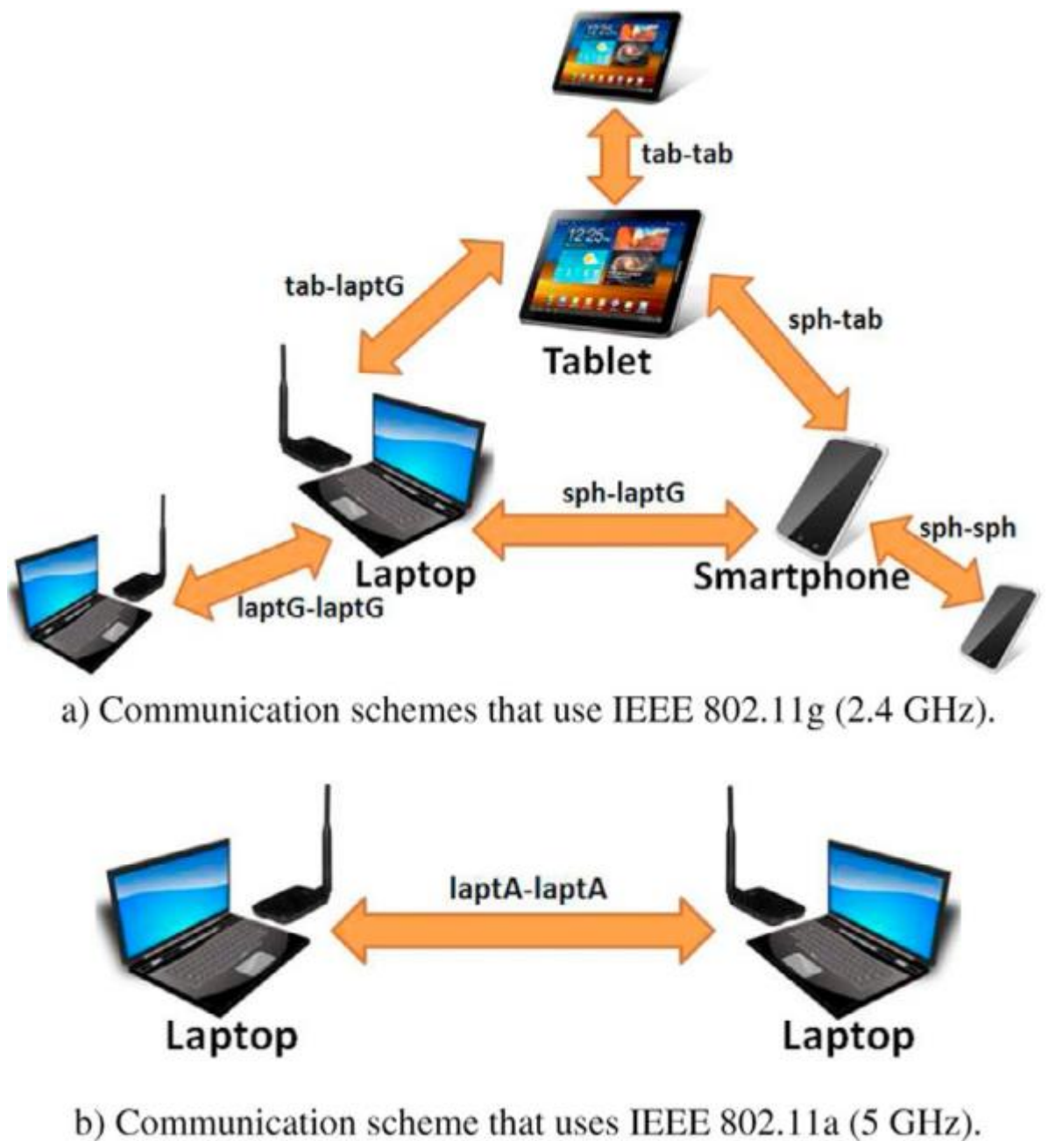

a) Communication schemes that use IEEE 802.11g (2.4 GHz).

b) Communication scheme that uses IEEE 802.11a (5 GHz).

**Fig. 8.** Representation of the communications analyzed in this study.

experiments consist in evaluating different QoS metrics of the VANET communications between two mobile vehicles.

Seven kinds of VANET communication schemes are analyzed in the *dynamic experiments*. These communication schemes are distinguished by the devices connected to exchange the information and the IEEE 802.11 standard. They are named using the following format: *<device1>* - *<device2>*. Note that for the laptops we have included a capital letter to specify the PHY/MAC standard used, i.e., "A" for IEEE 802.11a and "G" for IEEE 802.11g. The smartphones and the tablets always communicate using IEEE 802.11g, therefore, we have not used any letter to specify the standard. Thus the communication schemes analyzed here are: smartphone-smartphone (*sph-sph*), smartphone-tablet (*sph-tab*), smartphone-laptopG (*sph-laptG*), tablet-tablet (*tab-tab*), tablet-laptopG (*tab-laptG*), laptopG-laptopG (*laptG-laptG*), and laptopA-laptopA (*laptA-laptA*). Fig. 8 summarizes these seven communication schemes.

We have carried out 42 different tests for the *dynamic experiments*. Each test was defined according to one of the seven aforementioned communication schemes and one of the six distances between nodes (see Section 4.1). These tests are named according to the communication scheme and the distance: *<communication scheme>* - *<distance>*. For instance, *sph-sph-25 m* refers to the test in which both nodes are equipped with smartphones and are separated by 25 *m* while they are moving and *tab-laptG-125 m* represents the test in which one of the vehicles is equipped with a tablet and the other with a laptop (IEEE 802.11g) and they are separated by 125 m.

The V2V transferring data tests consist of exchanging data streams from the source node to the destination node of the VANET while the cars are moving. The information exchange is done by using ICMP (Internet Control Message Protocol) packets that encapsulate the bytes of data to be sent. We have selected this protocol as the use case protocol because we want to evaluate the feasibility of V2V (*peer-to-peer*) communications. We do not want to test any specific type of application or service, and therefore, we decided to avoid any connection-oriented protocol, such as the TCP (Transmission Control Protocol), or any other transaction-oriented protocol, as UDP (User Datagram Protocol). In order to study the effects of the size of data packets on the performance of the VANET communications, we performed data transfers by exchanging five kinds of data packets for each V2V test, that are defined by their size (32, 64, 128, 256, and 512 bytes of data). For each one of the five types of data packets, each VANET node transfers to the other streams of 100 independent packets each.

# 5. Experimental analysis

In this section, we present the different metrics that we have studied to evaluate the VANET communications and the results obtained after performing the real world experiments presented in Section 4.

## 5.1. VANET communication performance metrics

In the *static experiments*, the wireless transmission power of each device was evaluated in terms of the *received signal strength indicator* (RSSI) in a given location. RSSI reflects relative received signal strength in a wireless environment, in arbitrary units. Specifically, RSSI is an indication of the power level observed by a radio hardware while receiving a data frame. Remember that the evaluated RSSI includes the power from adjacent channel interference, thermal noise, etc. that could affect the signal received. In order to measure the RSSI we have used dBm (also known as dBmW), which is an abbreviation for the power ratio in decibels (dB) of the measured power referenced to one milliwatt (mW).

The feasibility of V2V communications evaluated during the *dynamic experiments* was analyzed in terms of three different QoS communication metrics:

- The *end-to-end delay* (E2ED), which is the time taken for a packet to reach the destination node (in milliseconds). This includes all possible delays introduced by intermediate nodes. Eq. 2 describes the E2ED incurred for any packet ($pi$) that has been correctly delivered.

$$E2ED_{pi} = arrive_time_{pi} - transmission_time_{pi} \tag{2}$$

- The *packet delivery ratio* (PDR), which refers to the ratio between the packets successfully delivered to the number of packets that have been sent by a source node (see 3).

$$PDR = \frac{number_of_packets_delivered}{number_of_packets_sent} \times 100\% \tag{3}$$

- The *transmission data rate* (TDR), which provides the amount of useful data successfully moved from one VANET node to another at a given time. In this study, the TDR is given in *kilobytes per second* (KB/s).

Now, we present the results on the four aforementioned metrics (transmission power plus QoS).

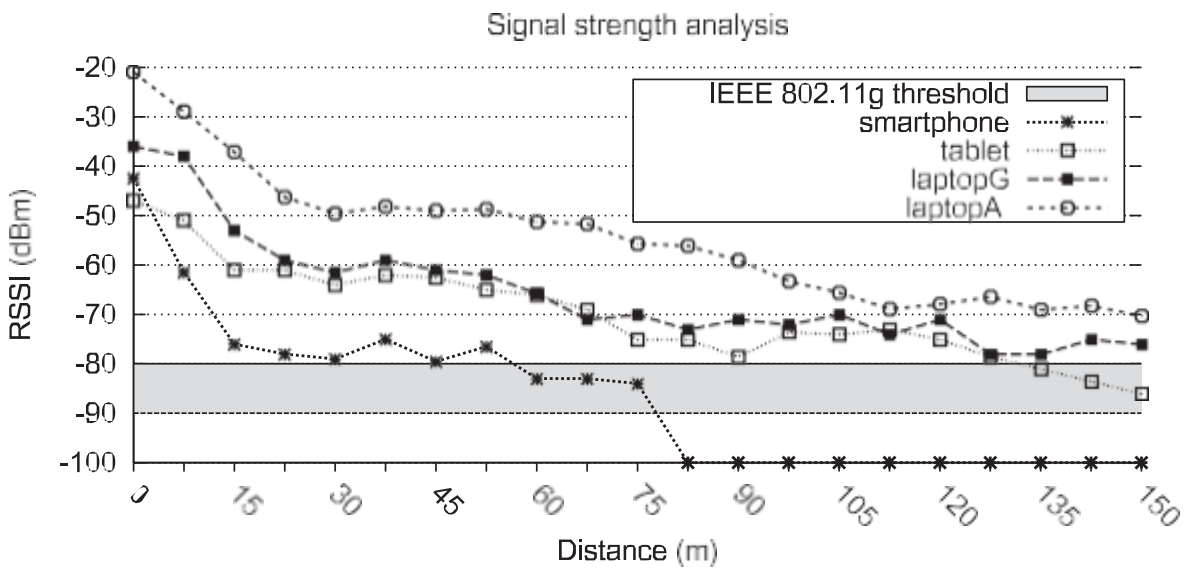

**Fig. 9.** Average signal strength results.

### 5.2. *VANET nodes transmission power results*

In general, the specifications of wireless interfaces given by the manufacturers provide values in ideal situations, therefore, we want to evaluate the Wi-Fi signal strength of the three devices studied here in real outdoor situations. As the experiments were carried out in the street on a normal road, the signal propagation was affected by different kinds of obstacles. Thus we offer the average of the signal strength sensed for 15 s.

The transmission power is important in wireless communications since it determines different aspects of the performance of the node in the network. Increasing the transmission power, not only increases the effective coverage and reduces the attenuation rate, but it also generates a greater amount of noise [32]. Thus, the most efficient transmission power is difficult to determine because of the combination of these positive and negative effects. However, empirical studies have demonstrated from an application point of view that the greater the signal strength received the better the PDR [33].

The experimental results are summarized in Fig. 9, which shows the average signal strength sensed at a given point for each studied device. The shaded area represents the lowest receiver sensitivities for which a receiver can correctly decode frames for most of IEEE 802.11g-based wireless devices (signal strengths between −90 dBm and −80 dBm).

As a first conclusion, the RSSI decreases with the distance between the node (device) that generates the signal and the measurement point (see Fig. 9). At the same time, we observe that, even if we have measured the signal strength for 15 s to limit the negative effects over the signal propagation in real outdoor scenarios, the values still show some irregular behavior. Our explanation for this phenomenon is that the RSSI includes the received power of the interference, thermal noise, etc. that has affected the possible regular behavior of such metric.

The smartphone is the least competitive device in terms of RSSI results. The signal strength values for this device are between −90 dBm and −80 dBm from 60 m to 75 m, which means that its performance in transmitting information could suffer from a degradation at these distances. After 75 m, the RSSI is lower than −90 dBm, and therefore, smartphones barely have the capacity to exchange data streams at greater distances.

The strength of the signal produced by the tablet is higher than −70 dBm for distances lower than 67.5 m. After that, from this point until 127.5 m, the strength of the signal decreases but it maintains values greater than −80 dBm. Thus, tablets may offer competitive communication performances at distances up to 127.5 m. Finally, when the distance is greater than 127.5 m the signal strength is lower than −80 dBm, but always higher than −90 dBm, which means that the tablet signal strength is over the lowest sensitivity threshold of IEEE 802.11g wireless based communications throughout our experiments.

The best results in terms of signal strength are provided by the laptop when using both PHY/MAC standards, IEEE 802.11a and IEEE 802.11g. On the one hand, when IEEE 802.11a is used, the received signal strength is significantly higher than in the other devices. The main reason is that the antenna gain is higher on the 5 GHz band than on the 2.4 GHz. In this case, the RSSI is lower than −60 dBm just when the distance is longer than 90 m and the lowest measured signal strength is −70.29 dBm (see Fig. 9). On the other hand, when the radio used in the laptop is configured with IEEE 802.11g, the RSSI results when the distance is lower than 67.5 m are surprisingly close to the ones obtained by the tablet. From this point until the furthest one the signal strength is always higher than −80 dBm.

Therefore, the best results in terms of transmission power are achieved by the laptop when transmitting on the 5 GHz band The second best are achieved when the same device uses the 2.4 GHz band. The third best RSSI is shown by the tablet, allowing interesting competitive results compared to the laptop for the complete experimentation (RSSI higher than −90 dBm). The lowest signal strength results are presented by the smartphone (as expected): so after 75 m the experimental results are lower than −90 dBm, offering an undesirable behavior. Applications needing more than 75 m would probably not rely on smartphone’s Wi-Fi, with the important exception of V2I, e.g., exchanging information to Wi-Fi spots and traffic lights in the city. We must also mention that, even if these results were expected, we are here quantifying the distance ranges and communication power of regular smartphones, something difficult to find in the related literature.

Finally, note that, although the signal strength decreases with the distance for all devices and frequency bands, the reduction is smoother and the signal strength is more stable when the radio used the 5 GHz band. This happens because the 2.4 GHz frequency band is way more crowded than the 5 GHz one, and therefore, the devices on the 2.4 GHz suffer much more interference than those on the 5 GHz.

### 5.3. *VANET communication feasibility experimental results*

In this section, we present the experimental results, analyzing the exchange of data between the two moving vehicles by evaluating the three metrics presented in Section 5.1: E2ED, PDR, and TDR.

The delivery delay results are shown in Table 1 in terms of the average E2ED in *milliseconds* (*ms*) of each data type and for each one of the seven V2V communication schemes analyzed in this study. The results are grouped by the distance between the cars during the data transfers. The cases in which none of the data packets sent reached the destination

**Table 1**
Experimental results in terms of the average V2V E2ED (milliseconds) grouped by distance between vehicles (separation) and the data packet type (size in bytes).

| | Connection | Data size in bytes | | | | | |
|---|---|---|---|---|---|---|---|
| | type | 32 | 64 | 128 | 256 | 512 | *aVg* |
| **Medium distance** | | | | ***Separation = 25 m*** | | | |
| | *sph-sph-25m* | 77.1 | 108.9 | 173.8 | 171.2 | 87.0 | 116.8 |
| | *sph-tab-25m* | 25.8 | 22.2 | 25.5 | 29.5 | 23.7 | 25.2 |
| | *sph-laptG-25m* | 18.6 | 8.8 | 7.7 | 8.8 | 9.1 | 10.0 |
| | *tab-tab-25m* | 30.0 | 20.9 | 71.6 | 47.7 | 51.7 | 40.6 |
| | *tab-laptG-25m* | 17.7 | 6.2 | 7.4 | 10.6 | 7.1 | 9.1 |
| | *laptG-laptG-25m* | 1.2 | 1.4 | 0.9 | 2.1 | 1.8 | 1.4 |
| | *laptA-laptA-25m* | 8.7 | 9.2 | 12.0 | 14.8 | 16.7 | 12.3 |
| | | | | ***Separation = 50 m*** | | | |
| | *sph-sph-50m* | 83.8 | 107.0 | 159.0 | 83.0 | 90.8 | 101.4 |
| | *sph-tab-50m* | 23.2 | 29.0 | 26.6 | 54.4 | 23.3 | 29.6 |
| | *sph-laptG-50m* | 29.2 | 9.0 | 9.2 | 21.2 | 21.9 | 16.2 |
| | *tab-tab-50m* | 27.7 | 51.5 | 40.2 | 38.8 | 126.0 | 48.9 |
| | *tab-laptG-50m* | 58.8 | 20.8 | 43.5 | 45.1 | 73.9 | 44.6 |
| | *laptG-laptG-50m* | 1.7 | 0.8 | 1.6 | 2.1 | 1.8 | 1.6 |
| | *laptA-laptA-50m* | 13.5 | 14.6 | 15.8 | 29.5 | 16.9 | 18.1 |
| | | | | ***Separation = 75 m*** | | | |
| | *sph-sph-75m* | 23.4 | 488.0 | 1042.0 | 158.0 | 187.0 | 203.8 |
| | *sph-tab-75m* | 32.3 | 26.4 | 36.1 | 41.3 | 51.7 | 36.6 |
| | *sph-laptG-75m* | 15.1 | 10.4 | 11.0 | 9.8 | 22.0 | 13.0 |
| | *tab-tab-75m* | 90.5 | 55.1 | 98.2 | 85.5 | 76.9 | 79.7 |
| | *tab-laptG-75m* | 33.2 | 8.9 | 7.8 | 10.5 | 8.7 | 11.6 |
| | *laptG-laptG-75m* | 2.2 | 1.6 | 2.5 | 2.1 | 2.2 | 2.1 |
| | *laptA-laptA-75m* | 14.6 | 19.6 | 14.6 | 15.5 | 17.7 | 16.5 |
| ***Long distance*** | | | | ***Separation = 100 m*** | | | |
| | *sph-sph-100m* | - | - | - | - | - | - |
| | *sph-tab-100m* | - | - | - | - | - | - |
| | *sph-laptG-100m* | 26.0 | 11.4 | 13.0 | 16.1 | 13.3 | 15.2 |
| | *tab-tab-100m* | 56.1 | 220.1 | 165.3 | 561.1 | 134.2 | 172.7 |
| | *tab-laptG-100m* | 66.7 | 55.2 | 78.5 | 149.7 | 26.3 | 64.7 |
| | *laptG-laptG-100m* | 2.8 | 3.2 | 7.0 | 5.7 | 6.9 | 4.8 |
| | *laptA-laptA-100m* | 96.2 | 146.0 | 154.9 | 207.1 | 168.7 | 154.6 |
| | | | | ***Separation = 125 m*** | | | |
| | *sph-sph-125m* | - | - | - | - | - | - |
| | *sph-tab-125m* | - | - | - | - | - | - |
| | *sph-laptG-125m* | 26.4 | 8.6 | 9.5 | 17.7 | 14.2 | 14.0 |
| | *tab-tab-125m* | 51.3 | 83.6 | 216.4 | 122.5 | 182.1 | 115.7 |
| | *tab-laptG-125m* | 27.2 | 23.3 | 10.2 | 7.5 | 13.5 | 14.6 |
| | *laptG-laptG-125m* | 1.6 | 1.7 | 1.5 | 2.0 | 2.7 | 1.9 |
| | *laptA-laptA-125m* | - | - | - | - | - | - |
| | | | | ***Separation = 150 m*** | | | |
| | *sph-sph-150m* | - | - | - | - | - | - |
| | *sph-tab-150m* | - | - | - | - | - | - |
| | *sph-laptG-150m* | 8.6 | 10.3 | 21.0 | 15.6 | 11.2 | 12.7 |
| | *tab-tab-150m* | 52.4 | 290.7 | - | - | - | - |
| | *tab-laptG-150m* | 132.0 | 175.1 | 78.2 | 73.2 | 151.0 | 114.8 |
| | *laptG-laptG-150m* | 2.0 | 1.8 | 1.7 | 3.6 | 4.9 | 2.6 |
| | *laptA-laptA-150m* | - | - | - | - | - | - |

node due to the loss of connection between the nodes are shown in the table by a dash (-).

In analyzing the E2ED results in Table 1, we have observed that the transmission times do not always grow with the node's distance or size of data packet transmitted. For instance, the average delay of *sph-sph-25m* (116 ms) is longer than that of *sph-sph-50m* (101.4 ms), even though the nodes were closer in the first case. Another example is the E2ED in *sph-laptG-75m* test, the delay exchanging packets with 32 bytes of data (15.1 ms) is longer than the ones transmitting bigger data packets.

Our explanation is that the experiments are carried out on an open road. Thus, as shown in Fig. 7.b, sometimes there are real world obstacles, i.e., different numbers of vehicles on the road. Also, wireless technologies sharing the same frequency band and operating in the same environment often interfere with each other, causing severe decreases in performance [34]. During our experiments, a number of other IEEE 802.11bg wireless networks appeared in our surroundings. All these issues negatively affecting the *ideal performance* of the wireless communications. This phenomenon is notable and reinforces the importance of this analysis: if we had been using a simulator these situations would not have been properly studied.

According to the results in Table 1, the communications that involve just smartphones spend the longest transmission times. Specifically, the longest average E2ED occurs when data is exchanged by vehicles equipped with smartphones separated by a distance of 75 m(*sph-sph-75m* test) and exchanging packets of 128 bytes (in this case the E2ED is 1042.0 ms). This time is longer than the times spent when smartphones exchanged packets of 256 and 512 bytes of data which is not an expected behavior. As we discussed above, this is due to the real world interactions negatively affect the *ideal performance* of the wireless communications. After 75 m, the smartphones lose the connection with each other. Remember that the RSSI of the smartphones after 75 m is lower than −90 dBm as it has been shown in the previous section.

The shortest transmission delays are observed when the packet exchanges are carried out between the vehicles equipped with laptops and wireless interfaces configured with the IEEE 802.11g standard. Indeed, the E2ED using such devices is between ten times and a hundred times shorter than the ones measured when the nodes are equipped with other kinds of devices. The average E2ED achieved during *laptG-laptG* tests is 2.2 ms.

Note that, although the RSSI results obtained by the laptop configured with IEEE 802.11a standard are more competitive than the ones obtained when using IEEE 802.11g, the *laptA-laptA* test E2ED delays are significantly longer than the *laptG-laptG* ones. The is principally because the performance of the network is more likely to be negatively affected by real world obstacles when it uses a higher frequency [35].

Table 2 summarizes the results of the entire experiment by showing the average and the relative standard deviation of the three metrics studied. The results are grouped by distances between the nodes (*medium distance* and *long distance* experiments). It also presents the global (average) results for the complete *dynamic experiments* set.

According to the results in Table 2, the communications in the testbed scenario in which both nodes are equipped with tablets (*tab-tab* tests) require longer E2ED than the scenario in which one node uses a smartphone and the other uses a tablet (*sph-tab* tests). This is not the expected behavior because the smartphone signal strength is lower than the one of the tablet (see Section 5.2), and therefore, our smartphone should show worse wireless communication capabilities than our tablets. The same occurs when we compare the E2ED of the *tab-laptG* and the *sph-laptG* tests. This can be explained by the road traffic density growth during the experiments with tablets (*tab-tab* and *tab-laptG*). Thus, the data transfers suffer from the existence of obstacles between the nodes.

Taking into account just the experiments carried out by using two laptops, we can observe that the average E2ED during the *laptA-laptA* tests is longer than ten times the E2ED during the *laptG-laptG* tests. This difference is significantly greater if we take into account just the average E2ED during the *long distance* tests (E2ED *laptA-laptA*=156.01 ms and *laptG-laptG*=3.06 ms). Thus, the IEEE 802.11g standard is more promising than the IEEE 802.11a to perform competitive vehicular communications.

The variability of the communications generated by the obstacles can be observed in the standard deviation of the E2ED results in Table 2. The largest deviation value, which means the lowest robustness, is obtained by the laptop communicating on the 5 GHz band. Moreover, as the road traffic increased during the experiments involving the tablets, the *tab-tab* and the *tab-laptG* transmissions presented larger deviation values than the *sph-tab* and the *sph-laptG* ones, respectively.

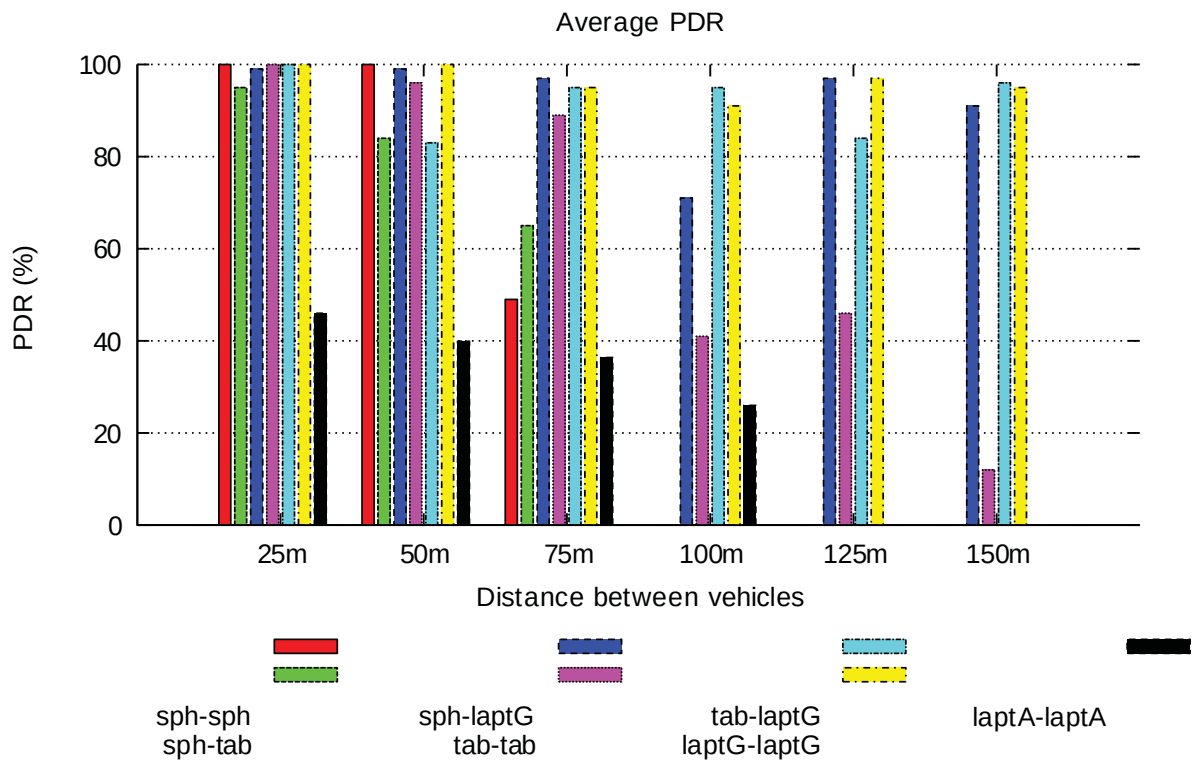

**Fig. 10.** Average PDR results.

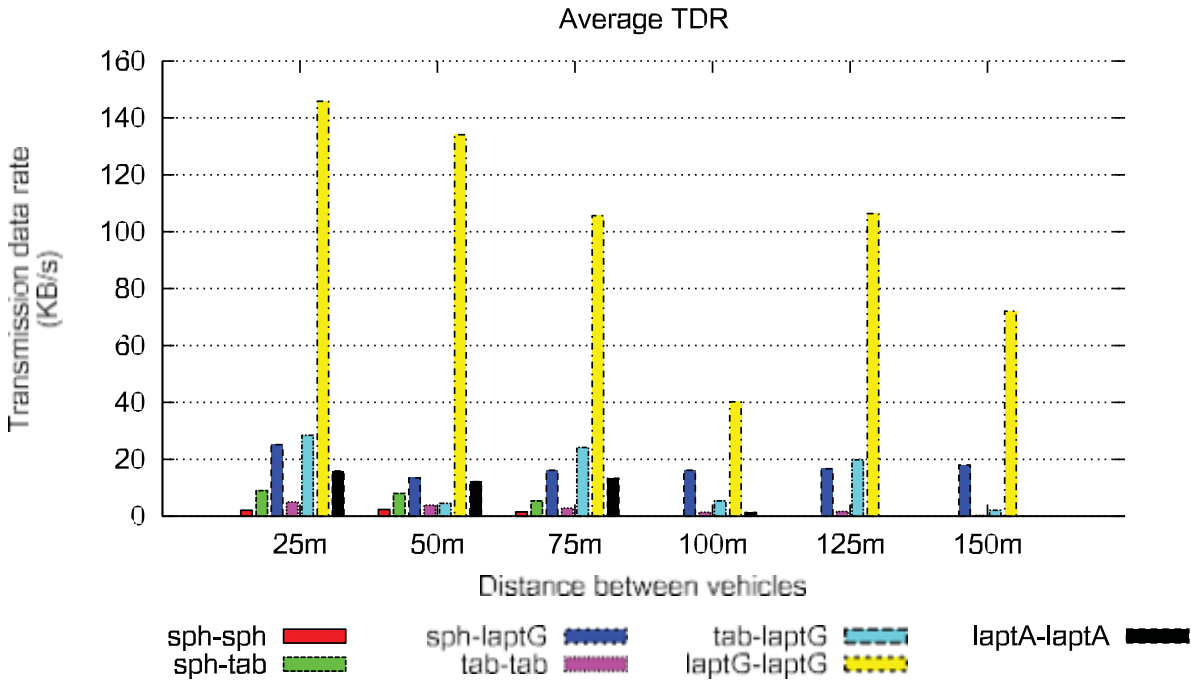

**Fig. 11.** Average TDR results.

Figs. 10 and 11 summarize the experimental results in terms of PDR and TDR, respectively, grouped by the distances between the VANET nodes. Note that, in both figures, some bars that represent the results of *sph-sph*, *sph-tab*, and *laptA-laptA* tests do not appear because there has been no data information exchange due to the complete loss of communication, as it can be seen in Table 2.

Analyzing the quantity of the successfully delivered data packets, we observe two clearly differentiated behaviors. On the one hand, all the VANET communications studied between devices that used the IEEE 802.11g standard (smartphones, tablets, and laptops) present similar and competitive results (PDR above 80%) for the scenarios in which the vehicles are separated by *medium distance* (see Table 2). However, only communications that involve laptops maintain such a high performance (PDR above 85%) when the distance passes beyond 100 m (*long distance*). On the other hand, when the vehicles exchanged data by using laptops configured with the IEEE 802.11a standard, the communications showed a limited performance since the PDR results were always below 50%. However, these nodes were not able to perform any data exchange when they were separated for distances greater than 100 m.

Taking into account just data transfers between smartphones (*sph-shp*), during the *medium distance* experiments,

**Table 2**
Average and relative standard deviation E2ED, PDR, and TDR results grouped by the distance between vehicles and the average for the complete experimentation.

| Connection type | E2ED (ms) | | PDR (%) | | TDR (KB/s) | |
|---|---|---|---|---|---|---|
| | Avg | Stdev(%) | Avg | Stdev(%) | Avg | Stdev(%) |
| | | | ***Medium distance (from 25 m to 75 m)*** | | | |
| *sph-sph* | 140.68 | 181.43 | 83.11 | 32.21 | 1.99 | 96.02 |
| *sph-tab* | 30.47 | 33.54 | 81.33 | 27.22 | 7.46 | 90.67 |
| *sph-laptG* | 13.06 | 51.68 | 98.45 | 2.82 | 18.26 | 80.90 |
| *tab-tab* | 56.43 | 52.82 | 94.89 | 7.14 | 3.91 | 69.69 |
| *tab-laptG* | 21.78 | 99.86 | 92.67 | 11.63 | 19.05 | 116.86 |
| *laptG-laptG* | 1.68 | 29.43 | 98.45 | 3.81 | 128.53 | 69.87 |
| *laptA-laptA* | 16.10 | 156.96 | 40.76 | 18.29 | 13.80 | 71.39 |
| | | | ***Long distance (from 100 m to 150 m)*** | | | |
| *sph-sph* | - | - | 0.00 | 0.00 | 0.00 | 0.00 |
| *sph-tab* | - | - | 0.00 | 0.00 | 0.00 | 0.00 |
| *sph-laptG* | 13.97 | 41.18 | 86.45 | 14.54 | 16.94 | 83.88 |
| *tab-tab* | 144.19 | 98.85 | 43.67 | 68.73 | 1.52 | 75.95 |
| *tab-laptG* | 64.71 | 87.39 | 91.67 | 8.44 | 9.11 | 146.81 |
| *laptG-laptG* | 3.06 | 63.29 | 94.45 | 8.08 | 72.86 | 67.42 |
| *laptA-laptA* | 156.01 | 63.80 | 8.67 | 157.57 | 0.44 | 192.93 |
| | | | ***Global average (from 25 m to 150 m)*** | | | |
| *sph-sph* | 101.33 | 251.88 | 41.56 | 111.12 | 0.99 | 167.79 |
| *sph-tab* | 15.70 | 65.10 | 40.67 | 108.52 | 3.73 | 161.92 |
| *sph-laptG* | 14.48 | 42.65 | 92.45 | 11.71 | 17.60 | 81.01 |
| *tab-tab* | 101.60 | 110.45 | 63.99 | 59.38 | 2.51 | 99.59 |
| *tab-laptG* | 47.59 | 101.79 | 92.17 | 10.02 | 14.08 | 133.09 |
| *laptG-laptG* | 2.51 | 63.59 | 96.45 | 6.48 | 100.71 | 76.01 |
| *laptA-laptA* | 38.42 | 259.38 | 24.71 | 79.21 | 7.12 | 135.65 |

the nodes successfully deliver 100% of the packets when the vehicles are separated by 25 m and 50 m, and then, its PDR decreases below 50% at 75 m (see *sph-sph* in Fig. 10) and, in the *long distance* ones, they do not exchange any packets at all (PDR=0%). Globally, the average PDR in this case is 45% with a relative deviation of 111%. In turn, the *sph-tab* experiments present similar PDR results (average PDR is 81.33% in *medium distance* and 0% in *long distance* experiments).

When both vehicles use tablets for the communications (*tab-tab* experiments), the PDR maintains a high performance in *medium distance*: 100.00%, 95.80%, and 88.66% when vehicles are separated by 25 m, 50 m, and 75 m, respectively. After that, when the distance between the vehicles is up to 125 m, they successfully exchange more than 40% of the packets. However, when the distance increases to 150 m, the PDR decreases to 12.20%.

Laptops transferred the highest amount of data between each other transmitting on the 2.4 GHz band (see PDR Table 2). The average PDR for the *laptG-laptG* communica- tions is 98.45% in *medium distance* and 94.45% in *long dis- tance*. In any case, the global average PDR is higher than 90% for all tests in which at least one of the communication nodes is equipped with a laptop and is using the IEEE 802.11g standard (see Table 2).

This is in sharp contrast to the results achieved by the same devices (laptops) when they communicate using IEEE 802.11a. The amount of delivered data are 40.76% and 8.67% for *medium distance* and *long distance*, respectively. In turn, the effective transmission range is shorter for the IEEE 802.11a communications than for the IEEE 802.11g ones. The main reason for this is that both standards use the same modulation (OFDM) and IEEE 802.11a transmits on higher frequencies (5 GHz over 2.4 GHz), reducing its communication capabilities against IEEE 802.11g [36,37].

In terms of transmission data rates, there are considerable differences between the *laptG-laptG* communications and the other VANET communications studied here (see Table 2 and Fig. 11).

Smartphones have the least competitive results in terms of TDR. In the *sph-sph* and *sph-tab* experiments, the nodes are able to exchange data in just *medium distance* scenarios by achieving average data rates of 1.99 KB/s and 7.46 KB/s, respectively. Although it seems a poor performance, the TDRs achieved for distances between 25 and 75 m are enough to deploy VANET applications for exchanging lightweight warning messages and traffic information with vehicles nearby and with the authorities via V2I.

When smartphones communicate with a node equipped with a laptop, the performance of the communications dramatically improves the TDR and the effective communication range. In *sph-laptG* experiments our nodes are able to exchange data in *medium distance* and *long distance* tests with an average TDR of 18.26 KB/s and 17.60 KB/s, respectively (see Table 2).

Tablets offer better TDR results than smartphones since *tab-tab* experiments achieve higher rates (its global average TDR 2.51 KB/s) and communicate with nodes at greater distances (up to 125 m). However, when one of the nodes is equipped with a laptop this does not hold true. The *sph-laptG* tests obtained more competitive TDR results than *tab-laptG* ones (global average TDRs, *sph-laptG*=17.60 KB/s and *tab-laptG*=14.08 KB/s). This is due to the increase in road traffic during the test that involved vehicles equipped with tablets, which negatively affected the communications.

Taking into account the experiments carried out by transferring data by using laptops configured with the IEEE 802.11a standard, the performance is significantly different for both *medium distance* and *long distance*. When the distance is shorter or equal to 75 m, these communications achieved a competitive transmission data rate (average TDR *laptA-laptA*=13.80 KB/s). However, the TDR drops to 0.44 KB/s for *long distance* tests.

As in the previously analyzed metrics, in the analysis of the data rates, the best results are achieved when the two VANET nodes use laptops using IEEE 802.11g to exchange data between each other. In the *medium distance* experiments the average *laptG-laptG* TDR is 128.53 KB/s and in the *long distance* ones the average TDR is 72.86 KB/s (see Table 2). These competitive transmission data rates may allow the exchange of multimedia information between the nodes, such as voice messages and videos. It is important to remark that this high TDR is achieved in distances up to 150 m.

Summarizing, the best TDR results are presented by communications using two laptops configured with the IEEE 802.11g standard (*laptG-laptG*). Then, the second and third best TDRs involved laptops too, *sph-laptG* and *tab-laptG*, respectively. These results clearly show that having a laptop in the communication loop makes smartphones and tablets usable over greater distances and in more difficult scenarios. This could be explained by the high sensitivity of the wireless interface used by the laptop. The other connections studied offered transmission rates between 2 KB/s and 8 KB/s.

Broadly speaking, globally analyzing the feasibility of the communications carried out on the 2.4 GHz band in terms of the three evaluated metrics (E2ED, PDR, and TDR), vehicles equipped with laptops offer a practical solution for deploying VANETs right away, because this solution allows communications at distances greater than 150 m with the largest PDR and TDR and the lowest E2ED. This is not an unexpected behavior because the laptops are equipped with the highest gain antenna that is placed outside the car (see Fig. 2.c), making the laptops communications free from the signal loss due to the required effort of signals to penetrate inside the car when tablets or smartphones are used. Analyzing the personal devices, both present competitive QoS and so could deploy useful VANETs with applications that do not require large transmission data rates. At the same time, tablets provided an effective communication coverage over 125 m, and smartphones a coverage up to 75 m. These relatively low performance results dramatically improve when devices with powerful wireless interfaces, such as laptops, are included in the VANET communication loop.

Note that, even the results obtained when the radios used the 5 GHz band are less competitive, the IEEE 802.11p-based communications are still considered as a promising technology for vehicular environments. This standard provides several improvements that make it more robust than IEEE 802.11a. An example of the changes described in its draft is the reduction of the bandwidth of the channels from 20 MHz of IEEE 802.11a to 10 MHz of IEEE 802.11p, which duplicates the transmission time for a specific data symbol, allowing the receiver to better cope with the characteristics of the radio channel in vehicular environments [38].

Finally, the importance of these outdoor testbeds is notable, because some of our experiments have suffered from the existence of real world obstacles producing a decrease in the performance of the wireless communications, leading to far from ideal results.

# 6. Conclusions

Widely available smartphones and tablets provide a set of facilities, which are required by VANET applications. Thus, they can be used to deploy VANETs even if specific on-board units are not available to most road users.

In this study, we have analyzed the use of smartphones, tablets, and laptops in vehicle-to-vehicle communications by defining a testbed carried out in the city of Málaga. In our analysis we have included two different MAC/PHY specifications: the widely used IEEE 802.11g and the IEEE 802.11a. We have designed two different types of experiments: the first designed to evaluate the wireless signal strength generated by the analyzed devices, and the second, focused on characterizing the QoS of V2V communications of these devices. In light of the analysis of our experimentation we can conclude that:

- In terms of signal strength, the best results were provided by the laptops with competitive RSSI for distances greater than 150 m. The best (the highest RSSI) is the laptop equipped with the radio using IEEE 802.11a and the second best is the same device utilizing IEEE 802.11g. The third ranked is the tablet with RSSI values are close to the ones obtained by the laptop transmitting on the 2.4 GHz band until the distance grew to 135 m. The least competitive device in terms of signal strength is the smartphone, but the RSSI maintained promising values for distances up to 75 m.
- According to the network feasibility obtained for the devices using the IEEE 802.11g standard, the smartphones allowed useful information to be exchanged with nearby nodes in urban areas. The tablets improved the smartphones' communication capabilities and could be used to exchange more information, also on highways. Finally, the laptops were able to exchange multimedia information (audio and video) with the highest data rates and with any kind of communication partner node (smartphones, tablets, and other laptops) at distances up to 150 m.
- The network feasibility when the laptops exchanged information with radios configured with the IEEE 802.11a standard were significantly less competitive than when they utilized IEEE 802.11g. They provided shorter effective communication range (up to 100 m), longer transmission delays, and lower transmission data rates.

It should be taken into account that the performance of communications has not always followed the regular expected behavior. This can be explained by the variability of the road traffic density (vehicles act as obstacles for the signal propagation) and by the existence of a number of other IEEE 802.11 wireless networks that interfered with the signal of our VANET. This kind of behavior, which influences the radio signal propagation, and therefore the communications QoS, is quite difficult to accurately represent and evaluate in simulators or emulators. Therefore, performing real world tests is strongly recommended to evaluate vehicular networks, even if it is difficult to master when conducting such experiments.

As a matter for future work, we plan to extend our testbed by including IEEE 802.11p experiments, as soon as we are able to find affordable devices on the market. At the same time, we are designing new outdoor experiments to evaluate VANETs performance, using more than two vehicles with different mobility models and taking into account different available routing and broadcasting approaches. Finally, we will define new testbeds by using other kinds of communication schemes such as V2I, in-vehicle, and cellular communications.

# Acknowledgments

J. Toutouh is supported by Grant AP2010-3108 of the Spanish Ministry of Education. This work is partially supported by the Regional Ministry for Economy, Innovation, Science, and Employment in Anadalusia and Innovation European FEDER, under UMA/FEDER FC14-TIC36 program. It is also partially funded by project number 8.06/5.47.4142 in collaboration with the VSB-Technical University of Ostrava. Universidad de Málaga, Campus de Excelencia Internacional Andalucía Tech.

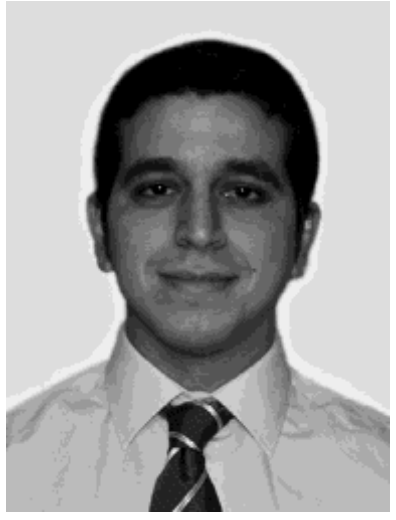

**Jamal Toutouh** is a Ph.D. candidate at University of Málaga (Spain). He received his M.S. degree in Software Engineering and Artificial Intelligence from the same university, M.S. degree in Information and Computer Sciences (Intelligent Systems speciality) at University of Luxembourg, and M.Eng. and B.Eng. degrees in Computer Sciences from University of Málaga. Jamal has published articles in three journals indexed by JCR Thomson ISI and in one other journal; one paper in LNCS and in other 13 referred international and national conferences. His current research focuses on the use of natural inspired algorithms (metaheuristics) to solve VANET design problems and to design new VANET applications and services. For this work, he has merited the Doctoral Consortium 2013 award given by the Spanish Association for Artificial Intelligence (AEPIA).

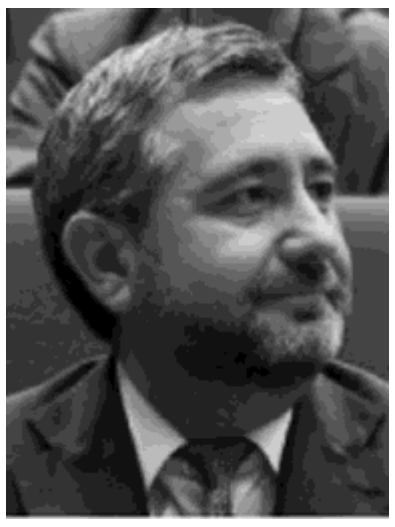

**Enrique Alba** had his degree in engineering and Ph.D. in Computer Science in 1992 and 1999, respectively, by the University of Málaga (Spain). He works as a Full Professor in this university with different teaching duties. Dr. Alba leads a team of several doctors and engineers (most of them PhD candidates) in the field of complex optimization. Dr. Alba has offered dozens doctorate courses and has directed several research projects. Also, he also works as invited professor at INRIA, the Univ. of Luxembourg, and Univ. of Ostrava. He is editor in several international journals and book series of Springer-Verlag and Wiley. He has published 81 articles in journals indexed by Thomson ISI, 17 articles in other journals, 40 papers in LNCS, and more than 250 refereed conferences. Besides that, Prof. Alba has published 11 books, 39 book chapters, and has merited 6 awards to his professional activities. Pr. Alba's H index is 42, with more than 8400 cites to his work.